# Waves on subwalength metallic surfaces: a microscopic view point

Philippe Lalanne[1] and Haitao Liu[2]

[1]Laboratoire Photonique, Numérique et Nanosciences, Institut d'Optique, Univ Bordeaux 1, CNRS, 33405 Talence cedex, France.
[2]Key Laboratory of Optical Information Science and Technology, Ministry of Education, Institute of Modern Optics, Nankai University, Tianjin 300071, China.

## 1. Introduction

In 1902, R.W. Wood observed that the spectrum of a continuous light source reflected by a metallic surface etched by a periodic array of tiny grooves exhibits a rapid variation that occurs within a range of wavelengths not greater than the distance between the sodium lines. Since then, grating anomalies have fascinated specialists of optics and physics, and nowadays with the progress of nanofabrication, metallic surfaces patterned with subwavelength indentations are studied for a variety of interesting properties, with applications ranging from sensing, new photonic and metamaterial devices, to integrated circuits mixing photonics and electronics [Ebb08]. It is the purpose of this chapter to examine the concepts, to elucidate the underlying physics and to discuss recent results and current problems in relation with resonance of metallic gratings. Particular emphasis will be placed on work carried out in the last decade, in anticipating future directions and in assessing the relevance of the subject to other areas of science.

Section 2 provides a historical background on the waves that are launched on a conducting surface by a subwavelength indentation, recalling the pioneer works at the beginning of last century by Marconi, Sommerfeld, Norton … in relation with long-distance radio-wave communications. Note that the field radiated by a dipole source in the vicinity of an interface has been considerably studied in the context of molecule fluorescence and other optical processes, such as surface-enhance Raman scattering, energy transfer … An interface may alter the way an excited molecule loses energy through fluorescence in several ways. It may absorb part of the spontaneous decay, and may alter both the radiative decay rate and the spatial distribution of the emitted radiation. Such situations are completely out of the scope of the present survey. The interested reader may refer to the review article [Bar98] and to other reviews quoted there.

At optical frequency, the field scattered by subwavelength indentations on metal surface has been first considered for understanding the spectacular Wood's anomalies. Well before the establishment of modern theories of gratings (Floquet-Bloch expansions, phenomenological models with the zeros and poles of scattering operators … ) and before the discovery of surface plasmon polaritons (SPP) by Ritchie [Rit57], microscopic theories of metallic gratings considered the SPP launched by the individual grating indentations as responsible for Wood anomalies. Section 3 summarizes the Fano's seminal ideas that, since 1941, have durably impacted the field of periodic metallic surfaces.

Section 4 describes what is presently known on the waves launched on metal surfaces by subwavelength indentations, which include the SPP mode and another contribution called the quasi-cylindrical wave (quasi-CW). A good knowledge of the properties of these waves is essential for understanding the rich physics of subwavelength metallic surfaces. Particular emphasis is put on 1D indentations such as ridges or grooves, 0D point defects such as holes being rapidly visited.

In Sections 5 and 6, we examine the scattering of SPPs and quasi-cylindrical waves by tiny indentations, emphasizing cross−conversion processes that convert quasi-CWs into SPPs and vice versa. Under the assumption that the indentations have subwavelength dimensions, scattering coefficients for the SPPs, quasi-CWs, and for a combination of theses waves can be consistently defined. The objective is to settle the foundation of a "microscopic" treatment of the electromagnetic properties of metallic subwavelength surfaces, which is accurate and intuitive. For that, the microscopic treatment should fit our current understanding and design recipes that all rely on a wavy description, which assume that surface waves are first generated by some illuminated indentations, then propagate on the metal surface and interact with nearby indentations before being recovered as freely propagating light or detected. Note the large gap existing between such an intuitive wavy picture and state-of-the-art numerical tools (some are purely numerical like finite−element and finite−difference methods, some are more physically oriented like modal− or multipole−expansion methods …), that rarely consider the waves launched on the surface and that never directly calculate how those waves are scattered by the subwavelength indentations. The main physical ingredients of our understanding of subwavelength surfaces (the launching, absorption, propagation and scattering of surface waves) are only implicitly taken into account in standard modelisation by matching the continuous electromagnetic field components at the interface.

Section 7 concludes the chapter.

## 2. Waves on metal surfaces: historical background

The field scattered by subwavelength indentations or emitted by subwavelength emitters in the vicinity of interfaces has been of longstanding interest in electromagnetism. In the 1900's, the rapid development of radio-wave technology prompted theoretical studies to explain why very long-distance (over-ocean transmission have been achieved in 1907 by Marconi) transmission could be achieved with radio waves above the earth. The solution is indeed linked to guiding by the ionosphere layers, but at the beginning of the 20th century, the explanation was thought to be due to the nature of the surface waves launched on the flat earth by the emitting antennas acting as a dipole. Sommerfeld was the first to determine the complete electromagnetic field radiated by a subwavelength antenna (a 0D vertical dipole) at the interface between two semi-infinite half spaces. He verified that his complicated solution [Som26] is composed of a "direct contribution" and of a bounded Zenneck mode [Zen07], the analogue of the surface plasmon polariton (SPP) [Rae88] for metals at optical frequencies, with an exponential damping. On the other hand, the amplitude of the direct contribution does not decay exponentially, but algebraically as $1/r^2$ at asymptotically long-distance from the antenna [Nor35, Nor37]. The direct contribution, known as the Norton wave [Col04, Bañ66], was therefore believed to be responsible for long-distance radio transmission.

In nanophotonics, the field scattered on metallic surfaces by subwavelength indentations is also essential, since it is responsible for the electromagnetic interaction between nearby indentations on the surfaces. Since the initial milestone interpretation of Wood's anomalies [Woo02] by U. Fano [Fan41] who introduced the concept of bounded SPP modes, SPPs have been central in modern history of the research on the optical properties of metallic surfaces, which have recently enabled researchers to overcome the diffraction limit for applications in microscopy [Spe92], nano-optical tweezing [Rig07], integrated optics [Dit02] and lasers [Zhe08]. From a mathematical point of view, the solution of this photonic problem is identical to that of the radio-wave problem [Lal09, Nik09]. However, there are also differences. We are mainly concerned by short-distance (rather than long-distance) electromagnetic interactions, since the distance between two neighboring indentations on subwavelength optical surfaces is of the order of $\lambda$ and rarely exceeds $10\lambda$. The second important difference concerns the fact that the dipole orientation cannot be chosen in nanophotonics. For instance, for a subwavelength 1D indentation under illumination of transverse-magnetic (TM) polarization, two coherent equivalent electrical dipoles of different polarizations are generally excited with different strengths.

Despite its importance for understanding the rich optics of subwavelength metallic surfaces, the field scattered by subwavelength indentations on a metal surface has been studied only recently. Lezec and his co-workers [Gay06] were the first to recognize the importance of a "direct" wave other than the SPP. This initial finding has been followed by theoretical [Lal06, Che06, Ung08, Nik10] and experimental [Aig07] works aiming at determining the main characteristics of this wave. It turns out that for intermediate distances of interest ($x < 10\lambda$), the direct wave is very different from the Norton wave; it looks like a cylindrical wave.

## 3. Fano's microscopic model of Wood anomaly

In 1902, R.W. Wood, when observing the spectrum of a continuous light source reflected by an optical metallic diffraction grating when the incident wave is polarized with its magnetic vector parallel to the grooves (TM polarization), noticed a surprising phenomenon: "*I was astounded to find that under certain conditions, the drop from maximum illumination to minimum, a drop certainly of from 10 to 1, occurred within a range of wavelengths not greater than the distance between the sodium lines*" [Woo02]. Wood's discovery drew immediately a considerable attention and the fascination of many specialists of optics for the so-called Wood's anomalies that never died.

By considering the metal as perfectly conducting and using a complicated mathematical derivation, Lord Rayleigh proposed the first explanation to the existence of the anomalies [Ray07]: an anomaly in a given spectrum occurs at a wavelength corresponding to the passing-off of a spectrum of higher order, in other words, at the wavelength given by the grating equation for which a scattered wave emerges tangentially to the grating surface. Considering the imprecise knowledge of the grating period in Wood's experiment, the agreement between the grating equation and Wood's experimental results was considered as rather fair, and the Rayleigh conjecture remained unquestioned during almost two decades. However, the conclusions radically changed in 1936, with Strong's study of Wood's anomalies for various metallic gratings having the same period [Str36]. Strong evidenced that the anomalies occur at a wavelength systematically larger than that predicted by the grating equation.

To explain the red shift from the Rayleigh condition, U. Fano introduced a microscopic model of Wood's anomalies in his seminal article published in 1941 [Fan41] (40 years after Wood's observation). Fano's model is much less mathematically involved than the theoretical work by Lord Rayleigh. It rather relied on a Huygens-type very intuitive interpretation, and importantly, it suggested that a surface mode with a parallel momentum greater than the free space momentum be involved in the energy transport between adjacent grooves. It is retrospectively interesting and amazing to see how the surface wave, which is nothing else than the SPP of the flat interface, is introduced in Fano's model. U. Fano first considered the parallel propagation constants of the modes of a glass plate sandwiched between a metal and a vacuum and asks himself "*Is there left any mode when the thickness of the glass layer vanishes?*". By solving analytically the bi-interface problem, he showed that one and only one bound mode (the SPP) exists in the limit of vanishingly small glass thicknesses for TM polarization, with a complex propagation constant whose real part is always slightly larger than the modulus $k_0$ of the wave-vector in a vacuum. He therefore made the ansatz that Wood's anomaly originates from a collective resonance of the subwavelength surface (see Fig. 1), in which the part of the wave scattered by groove *A* excites the bound mode, travelling along the surface with a phase velocity smaller than the vacuum phase velocity, which gives a resonance whenever it reaches the neighboring groove *B* in phase with the incident wave (phase-matching condition). Denoting by $k_{SP}$ (surprisingly Fano does not give any analytical expression) the complex propagation constant of the surface wave and assuming that the grooves are infinitely small and thus neglecting multiple scattering, the microscopic interpretation by Fano leads to the following phase matching condition,

$$\text{Re}(k_{SP}) = k_x + 2\pi/a, \qquad (1)$$

where the real part of the propagation constant is matched to the parallel wave vector $k_x$ of the incident plane wave through a wave vector $2\pi/a$ of the 1D reciprocal lattice associated to the grating (*a* being the periodicity). In Rayleigh's theory, because perfect metals were considered, the wave on the perfectly-conducting surface propagates exactly with the vacuum phase velocity, and this causes the phase velocity difference that explains the red-shift for real metals.

The big success of introducing a bounded SPP mode to convincingly explain the experimental red shift is a milestone result, and since Fano's work our understanding of Wood's anomalies is intimately linked to the resonant excitation of SPPs. The last decade has proved that this vision is simplistic and that, although Fano's approach is remarkable in predicting the resonant wavelength, Rayleigh and Fano's interpretations should be actually "combined" to provide a quantitative analysis of Wood's anomalies.

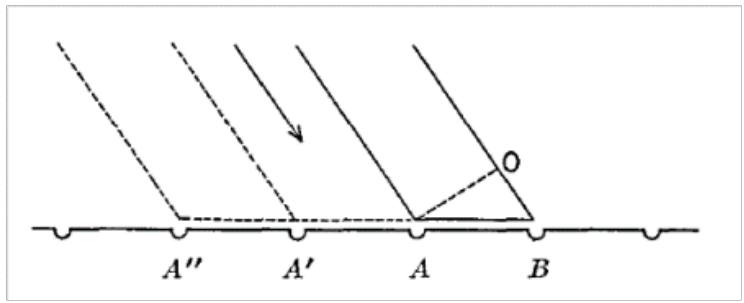

**Fig. 1. Fano's microscopic model of Wood's anomaly (from [Fan41]).** In Rayleigh's interpretation derived by considering the metal as a perfect conductor, resonance occurs whenever the part of the wave that is scattered by groove *A* and that is traveling along the grating with the vacuum phase velocity reaches the neighboring groove *B* in phase with the incident wave and with the waves scattered by the grooves *A'*, *A"*. What Fano proposes to explain the red-shifted Wood anomaly is to replace the free-space grazing wave of Rayleigh by a bounded mode with a smaller phase velocity. This bounded mode is nothing else than the SPP of the flat metallic surface, which will be discovered 16 years after by Ritchie [Rit57].

## 4. Field scattered on metal surfaces by subwavelength indentations

The waves scattered on the surface of a metal film by a tiny indentation are at the essence of the optical properties of metallic subwavelength surfaces. Provided that the indention is small enough (in practice, dimensions should be smaller than λ/3), it is convenient to consider asymptotically small indentations. For 1D indentations, such as grooves or ridges like in Wood's experiment, the equivalent sources are Dirac line sources. When the polarization of the incident wave is parallel to the indentation, the field radiated on the surface is nearly zero. In that case often referred to as TE polarization, the set of indentations on a metal surface are all independent and the field scattered by all indentations is simply the superposition of the field scattered by every indentation. This "trivial" case is not discussed hereafter. The TM polarization case is much more interesting. Two coherent electric line sources, one being polarized perpendicular to the surface and the second one being polarized parallel to the surface and perpendicular to the indentation, have to be considered. This case is discussed in the next subsection. For 0D indentations such as holes, three dipole polarizations should be investigated. The properties of the associated radiated fields will be qualitatively discussed in the second subsection.

### 4.1 The quasi-cylindrical wave

The scattering of a 1D subwavelength indentation illuminated by a TM wave has been the subject of intense research [Gay06, Lal06, Che06, Aig07, Dai09, Ung08] over the last decade. Hereafter we simply summarize the main results, which are documented in a review article [Lal09].

Referring to Fig. 2, a subwavelength indentation invariant along the *y*-axis (the *z*-axis being perpendicular to the surface) and illuminated with a plane wave polarized in the *x-z* plane (Fig. 2a), can be replaced by two electric line sources in the dipolar approximation (Fig. 2b), one $J_z$ being polarized perpendicularly to the interface (along the *z*-axis) and the other one $J_x$ parallel to the interface (along the *x*-axis). Concerning the field scattered on the surface (this is the field that is responsible for the electromagnetic interaction between the indentations on the surface), three important properties are worth mentioning here.

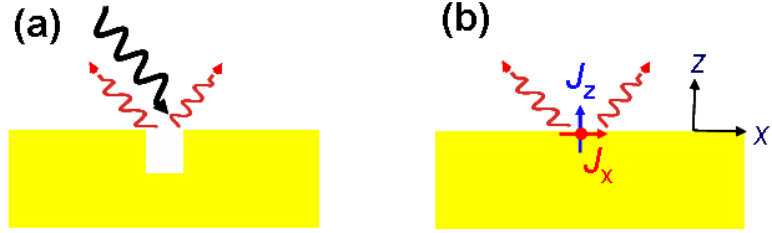

**Fig. 2. Equivalence of a subwavelength indentation under TM illumination (a) to two electric line sources (b).**

**Property 1:** *The field radiated on the surface by each individual line source can be decomposed into a SPP mode and a quasi-cylindrical wave (quasi-CW), which represents a "direct" contribution from the source.*

At optical frequencies, the amplitude of the quasi-cylindrical wave is initially damping as $x^{-1/2}$ (just as a cylindrical wave) in the vicinity of the line source, then is dropping at a faster rate for intermediate distances $\lambda < x < 10\lambda$, before reaching an asymptotic regime behavior with an $x^{-3/2}$ damping rate at large propagation distances.

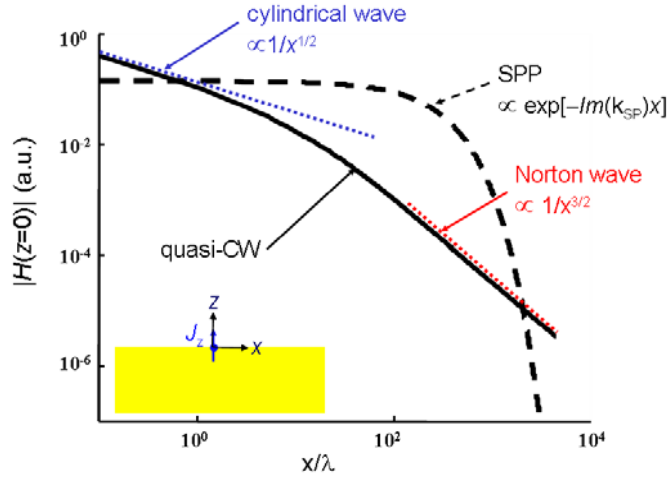

**Fig. 3. Magnetic field radiated at λ=1 μm on an air/gold interface ($z = 0$) by a line source $J_z$ polarized vertically.** The field is composed of a SPP (dashed curve) and of a quasi-CW (solid curve). The latter takes two asymptotic forms. It is very intense and behaves as a cylindrical wave (dotted blue line) with a $1/x^{1/2}$ decay rate at small propagation distances. At very long propagation distances, it is very weak and decays as $1/x^{3/2}$. It is the analogue of the Norton wave (shown with the dotted red line) discovered for radio communication.

Figure 3 illustrates the different contributions to the magnetic field radiated on an air/gold (permittivities $\varepsilon_d = 1$ and $\varepsilon_m = -46.8 + 3.5i$) interface ($z = 0$) by a line source vertically polarized. The results hold for gold at $\lambda=1$ μm. The dashed line is the SPP contribution, with an exponential damping $\exp[-Im(k_{SP})x]$, and the solid curve is the "direct" wave contribution. At very large propagation distances, the direct-wave decay rate asymptotically tends to $1/x^{3/2}$ and becomes the analogue of the Norton radio wave (shown with the dotted red line) at optical frequencies. For subwavelength propagation distances ($x < 2\lambda$), the direct wave contribution dominates. It is very different from the Norton wave as it looks like a cylindrical wave with a $1/x^{1/2}$ damping rate (dotted blue line). Consistently, the direct wave contribution has been called a quasi-cylindrical wave (quasi-CW) in the recent literature. The existence and importance of the quasi-cylindrical wave at optical frequencies on metals has been first observed with a very elegant slit-groove experiment [Gay06], in which the groove acts as a line source and the slit as a local detector of the field scattered by the groove. By systematically varying the groove-slit separation-distance in a series of samples, the field pattern is recorded. The experimental data, which were probably contaminated by an undesired adlayer on the silver film, have been initially interpreted in a confusing manner as shown in [Lal06, Aig07], but they had the merit to unambiguously reveal the existence and importance of a direct wave (different from the SPP) that is initially dominant for $|x| < 2\,\lambda$.

**Property 2:** *As one moves from the visible to longer wavelengths, the SPP is less attenuated, but it is also less and less efficiently excited, whereas the quasi-CWs are equally excited at all energies.*

This property is illustrated in Fig. 4 for different wavelengths. More precisely, the figure represents the magnitude of the total magnetic field $H(x, z = 0)$ radiated on the interface ($z = 0$) by a $z$-polarized line source located at $x = z = 0$. The total field results from the sum of two contributions, $H(x) = H_{SP}(x) + H_{CW}(x)$, where $H_{SP}$ (blue-dotted) and $H_{CW}$ (red-solid) represent the SPP and quasi-CW contributions, respectively. We first note that the initial quasi-CW contribution at short distances is nearly independent of the metal dielectric properties, whereas the initial SPP contribution rapidly drops as the metal conductivity increases, $|H_{SP}| \propto |\varepsilon_m|^{-1/2}$. At visible wavelengths ($\lambda = 0.633\,\mu m$), the SPP contribution dominates even at relatively short distances, the SPP and quasi-CW being actually equal for $x_c \approx \lambda/6$. At thermal-infrared wavelengths ($\lambda = 9\,\mu m$), the quasi-CW is preponderant until distances as large as $100\lambda$. It can be shown that the initial crossing distance $x_c$ below which the quasi-CW wave dominates increases with the metal conductivity, $x_c \approx \lambda|\varepsilon_m|/(2\pi\varepsilon_d^{3/2})$ [Lal09].

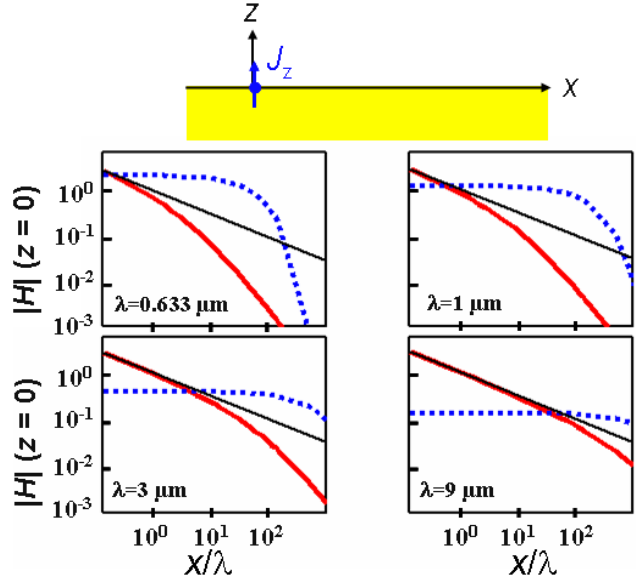

**Fig. 4. Magnetic field, $H(x) = H_{SP}(x) + H_{CW}(x)$, radiated by a vertically-polarized line source $J_z$ at an Ag/air interface (inset on the top) for wavelengths ranging from the visible to thermal infrared (from [Lal06]). The blue dashed curves correspond to $|H_{SP}|$ and the red-solid curves to $|H_{CW}|$. Thin black lines show a damping scaling as $1/x^{1/2}$. The calculations are performed**

for silver but similar results have been obtained for gold. Note the logarithmic scales used in both the horizontal and vertical axes, which are all identical for the sake of comparison. The frequency-dependent value of Ag permittivity is taken from [Pal85].

**Property 3:** *The quasi-cylindrical waves radiated on the surface by each individual line source, $J_x$ or $J_z$, although they differ in amplitude and phase, are almost identical in shape.*

Figure 5 illustrates the property. In the left graph, the magnetic fields of the quasi-cylindical waves radiated by vertical (blue curve) and horizontal (red curve) line sources ($J_x = J_z = 1$) are shown as a function of the distance $x$ from the source. The calculation is performed for a gold substrate at $\lambda = 800$ nm. In the right panel, the source $J_x$ has been optimized ($J_x \approx 3i$) so that its associated quasi-cylindrical wave is similar to that generated by the vertical source. It turns out that, although a slight difference remains, the two fields are almost superimposed. It can be shown that this difference becomes smaller and smaller as the metal conductivity increases (or $\lambda$ increases) [Lal09].

This property has an important consequence if one neglects the small residual difference. When a subwavelength indentation is illuminated by a TM polarized light, the scattered field can be seen as the total field radiated by two line sources, $J_x$ and $J_z$, and the relative amplitudes of the line sources are arbitrary: they for instance depend on the incident illumination (its angle of incidence for instance if it is a plane wave), on the actual geometry of the indentation, on the dielectric and metal permittivities… A priori two independent radiation problems should be considered, but since the quasi-cylindrical waves associated to the two line source polarizations are identical in shape, any arbitrary sub-$\lambda$ indentation illuminated by any incident electromagnetic field will launch a unique field (the quasi-cylindrical wave) on a metallic surface, in addition to the SPP. Also note that the mixing ratio between the SPP and the quasi-CW radiated by the two line sources are approximately the same [Lal09], and we finally conclude that this mixing ratio is fixed for the field scattered by any sub-$\lambda$ indentation.

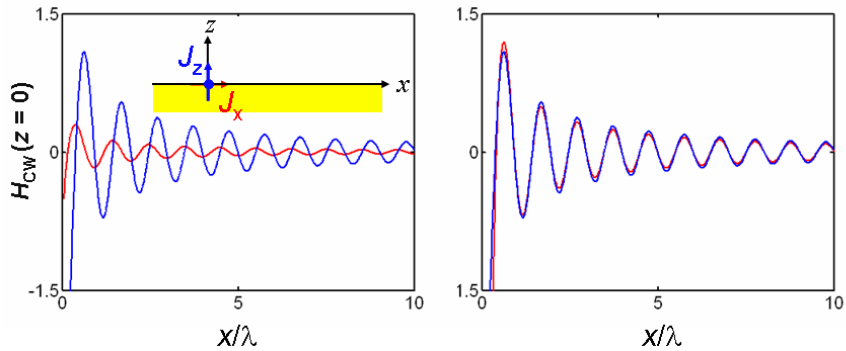

**Fig. 5. Illustration of property 3 for a gold/air interface at $\lambda = 800$ nm.** Left: The blue and red curves represent the magnetic field of the quasi-cylindrical waves radiated on the surface by two line sources polarized vertically and horizontally, respectively, with $J_x = J_z = 1$. The two quasi-cylindrical waves seem completely different a priori. Right: In reality, the two quasi-cylindrical waves are almost identical in shape and only differ by a constant, as show by the new red curves obtained for $J_x \approx 3i$. The frequency-dependent Au permittivity takes value from [Pal85].

## 4.2 SPP and quasi-CW launched by a dipole point-source on a metal surface (3D case)

The field radiated by point-source or sub-λ antennas in the vicinity of metallic surfaces has been of long-standing interest in classical electromagnetism. In particular for long-distance radio propagation and for remote sensing, the problem was analyzed in detail by Sommerfeld [Som09,Som26], Norton [Nor35,Nor37] and others for a half-space conductor with a finite conductivity (the sea surface for instance). The conclusions were that the radiated field can be calculated as an integral along a contour in the complex plane and is composed of two contributions, the Zenneck mode (corresponding to the pole, the analog of the SPP at visible frequencies) and a "direct" wave (corresponding to branch integrals, the analog of the quasi-CW). In an intermediate region and near the surface, the field is well approximated by that of the cylindrical Zenneck mode; but then, as the distance increases further, the long-distance propagation is mainly due to the direct wave that is often referred to as the Norton wave. The latter, whose amplitude asymptotically decays as $1/r^2$, overcomes the Zenneck mode at large distances because of the additional exponential damping factor $\exp(ik_{SP}r)$ of the Zenneck mode. These issues are discussed in great detail in the review article by R.E. Collin [Col04] or in the book by Baños [Bañ66].

Hereafter we simply show an example for the sake of illustration. For a vertical dipole, perpendicular to the interface, the in-plane component of the radiated field is radially polarized and isotropic. The situation is more interesting for an in-plane dipole (let us say parallel to the $x$-axis). Both the SPP and quasi-CW fields on the surface are anisotropic. Figure 6 shows the radial electric fields radiated by such a dipole. The SPP field is proportional to $r^{-1/2}\exp(ik_{SP}r)$ and its electric vector is mainly perpendicular to the surface with a small in-plane component. Along any direction (different from $\theta=\pi/2$) the in-plane SPP field tends to be radially polarized, as $E_\theta/E_r \approx \tan(\theta)/r$, where the subscript $\theta$ and $r$ are used to denote the orthoradial and radial components of the fields. The quasi-CW wave initially varies as $r^{-1}\exp(ik_0r)$ at small distances from the dipole, then at longer distances, its electric field amplitude decays algebraically with distance as $r^{-2}\exp(ik_0r)$, like the Norton radio wave. The electric field of the quasi-CW points mainly along the direction perpendicular to the interface, and its in-plane components satisfy $E_\theta/E_r \approx \tan(\theta)$. However, for the vertical dipole the in-plane component of the quasi-CW field is radially polarized and isotropic.

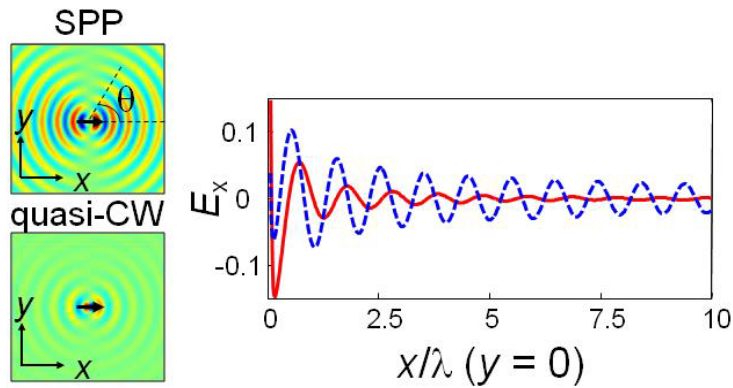

**Fig. 6. Radial electric field radiated on an Au/air surface by a dipole point source polarized parallel to the surface (along *x*-axis) for λ=800 nm.** The left panels show a surface map for the SPP (top) and quasi-CW (bottom) fields. The same units are used in both plots. As the metal

conductivity increases, the perfect-conductor limit is reached and the quasi-cylindrical wave becomes a spherical wave with a $r^{-2}\exp(ik_0r)$ behavior. In the right panel, the radial field ($E_x$) is plotted as a function of $x$ for $y = 0$. The blue dashed curve corresponds to the SPP mode and the red-solid curve to the quasi-CW. The frequency-dependent value of Au permittivity is taken from [Pal85].

## 5. Application of the Fano model to the extraordinary optical transmission (EOT)

The modes (SPP) and waves (the quasi-CW) scattered by individual subwavelength scatterers are at the essence of the electromagnetic properties of subwavelength metallic surfaces and in particular of subwavelength gratings. Every individual indentation that is illuminated launches SPPs and quasi-CWs on the surface. The launched fields further interact with the adjacent indentations, before being eventually radiated back into free space.

The electromagnetic interaction on the surface may lead to a complicated multiple scattering process, in which for instance the launched quasi-cylindrical waves may scatter and generate SPPs, or vice-versa. Before examining the multiple scattering process in details in Section 6, to simplify we will examine a simplified model, assuming like Fano, that the electromagnetic interaction among the indentations is only mediated by the SPPs of the flat interfaces between the indentations, the quasi-CW contribution being neglected. There are two reasons for considering such a pure-SPP model that only considers SPPs. First the model allows us to define the SPP scattering coefficients for the individual indentations, and these scattering coefficients are fundamental to understand the multiple scattering process (since the same scattering coefficients apply to the quasi-CWs as well, as shown in Section 6). Second, by comparing the predictions of the model with fully-vectorial computational results, one may directly determine the role of SPP in the electromagnetic property of subwavelength surfaces [Liu08].

To illustrate our purpose, we will consider the text-book case of the extraordinary optical transmission (EOT). The EOT was first observed in the near infrared with subwavelength hole arrays perforated in opaque gold and silver films [Ebb08], and is an emblematic example in plasmonics that has sparked a huge amount of research trying to apply the phenomenon and to unveil the underlying mechanisms, and especially to unveil the role of SPPs in the transmission. The analysis is performed for a self-supported membrane (thickness $d$) in air for the sake of simplicity (the upper and lower grating interfaces are identical, see Fig. 7a).

At a microscopic level, the basic mechanism enabling the EOT is a coherent diffraction by all the individual holes acting as elementary scatterers. However, it is more convenient to consider isolated 1D arrays of holes (a periodic hole chain with periodicity $a$ in the $y$-direction, see the bottom panels in Fig. 7) perforated in a metal substrate as the elementary scatterers. Provided that the hole separation distance is subwavelength, the 1D hole chains act as 1D indentations, like in classical metallic gratings.

The elementary SPP-scattering events used in a pure-SPP model of the EOT are shown in Fig. 7b-d for classical diffraction geometries (the $y$-component $k_y$ of the in-plane wave vector momentum is zero). Upon interaction with the chain, the SPP modes are partly excited, transmitted, reflected or scattered into the chain mode and into a continuum of outgoing plane waves. The interaction defines four elementary SPP scattering coefficients. Two coefficients, see Fig. 7b, namely the SPP modal reflection and transmission coefficients, $\rho_{SP}$ and $\tau_{SP}$, correspond to in-plane scattering. The other two, $\alpha_{SP}$ and $\beta_{SP}$, correspond to the transformation of the SPPs into aperture modes or radiation waves, and vice versa. They allow us to link the local field on the surface to the far field that is transporting light away from the metal film.

From these elementary SPP scattering coefficients, a coupled-mode model that provides closed-form expressions for the transmittance and reflectance coefficients of the fundamental supermode of the 2D hole array, $t_A$ and $r_A$, is readily derived [Liu08]. For instance, the reflection coefficient $r_A$ of the fundamental supermode, a very important physical quantity of the EOT phenomenon [Mar01], can be written

$$r_A = r + \frac{2\alpha_{SP}^2}{u^{-1} - (\rho_{SP} + \tau_{SP})}. \qquad (2)$$

In Eq. (2) that holds for normal incidence ($k_x=0$), $u=\exp(ik_{SP}a)$ is the phase delay accumulated by the SPP over a grating period and $r$ is the reflection coefficient of the fundamental mode of the hole chain, see Fig. 7c.

It is crucial to realize that the SPP scattering coefficients in Eq. (2) are not related to the periodicity of the structure and that the coupled-mode model can be applied to aperiodic structures as well [Liu08]. Indeed, the essence of Eq. (2), and in particular of the denominator that results from a geometric summation over all chain contributions, is a multiple scattering process that involves the excitation of SPP modes by the incident field and the further scatterings of the excited SPPs onto the infinite set of periodically-spaced hole chains. The same denominator would be obtained for the groove geometry in Fig. 1, and would explain the Wood anomaly in Fano's interpretation.

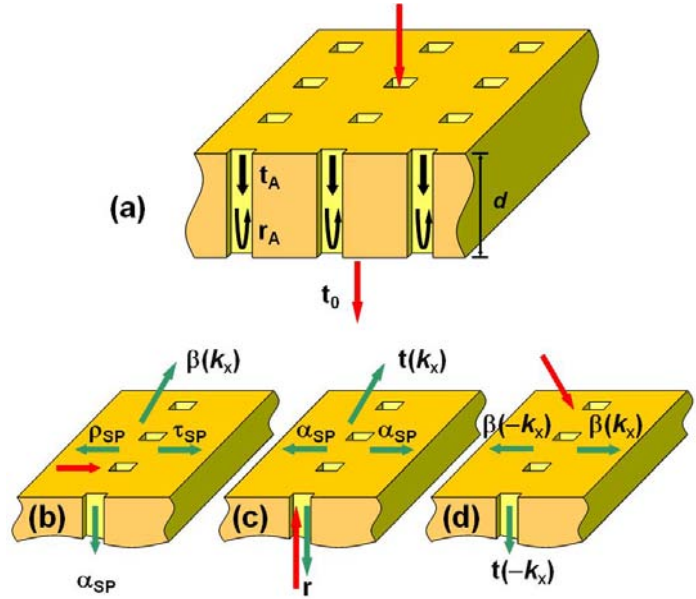

**Fig. 7. Pure-SPP model of the EOT.** (a) Self-supported geometry in air considered for the sake of simplicity. The transmission coefficient of the membrane (from the incident plane wave to the (0,0)th-order transmitted plane wave) is denoted by $t_0$. Similarly we denote by $t_A$ and $r_A$ the transmission and reflection coefficients of the fundamental supermode of the 2D hole array. The membrane thickness is denoted by $d$. **(b)-(d)** SPP elementary scattering processes involved in the

EOT. They are all associated to a single 1D hole chain of infinite depth under illumination by (**b**) the SPP mode, (**c**) the fundamental supermode of the hole chain, and (**d**) an incident TM-polarized (magnetic vector along the chain direction) plane wave impinging at an oblique incidence defined by its in-plane wave-vector component $k_x$. The red and green arrows refer to the incident and scattered modes, respectively. The processes in (**b**)-(**d**) define six independent elementary scattering coefficients, $\rho_{SP}$ (reflection coefficient of the SPP mode), $\tau_{SP}$ (transmission coefficient of the SPP mode), $\alpha_{SP}$ (scattering coefficient from the SPP mode to the fundamental supermode and vice versa according to the reciprocity theorem), $\beta(k_x)$ (scattering coefficient from the SPP mode to the outgoing plane wave with an in-plane wave-vector component $k_x$ and vice versa), $t(k_x)$ (scattering coefficient from the fundamental supermode to the plane wave and vice versa) and $r$ (reflection coefficient of the fundamental supermode).

The question arises on how accurate is the pure-SPP model in predicting the EOT phenomenon. The answer is provided in Fig. 8, which compares the pure-SPP model predictions (blue dashed curves) with fully-vectorial computational results (red solid curves). The comparison is performed for three spectral intervals, from the visible ($a$=0.68 µm) to the near-infrared ($a$=2.92 µm). The SPP model quantitatively predicts all the salient features of the EOT, and especially the Fano-type spectral profile with the antiresonance transmission dip followed by the resonance peak. Importantly, there are also some discrepancies that are due to the model assumption of a pure SPP electromagnetic interaction between the hole chains. As deduced from Fig. 8, the SPPs account for only half of the total transmitted energy at peak transmittance at visible frequencies, and only one fifth at longer wavelength in the near-infrared. The reason comes from the presence of the quasi-CW, which becomes more and more predominant as the wavelength increases, see Fig. 4.

This theoretical prediction has been recently confirmed experimentally by measuring the transmissions of a set of metal hole arrays with varying hole densities. More specifically, Beijnum and his coworkers have varied the size of the unit cell along the *x*-axis, choosing $a_x = qa\_(a = 450$ nm) and $a_y = a$, where $q$ is an integer ranging from 1 to 7 [Bei12]. When the measured transmissions are rescaled to correct for the reduced density of holes, all the arrays, $q$=2-7, except the $q$=1 array exhibit almost identical transmission spectra. Remarkably, all those rescaled spectra are reproduced with high accuracy by the pure-SPP model. In comparison, the $q$=1 array differs by a two-fold increase of the scaled transmission peaks, a distinct effect that is attributed to the impact of the short-range-interaction provided by the quasi-CW.

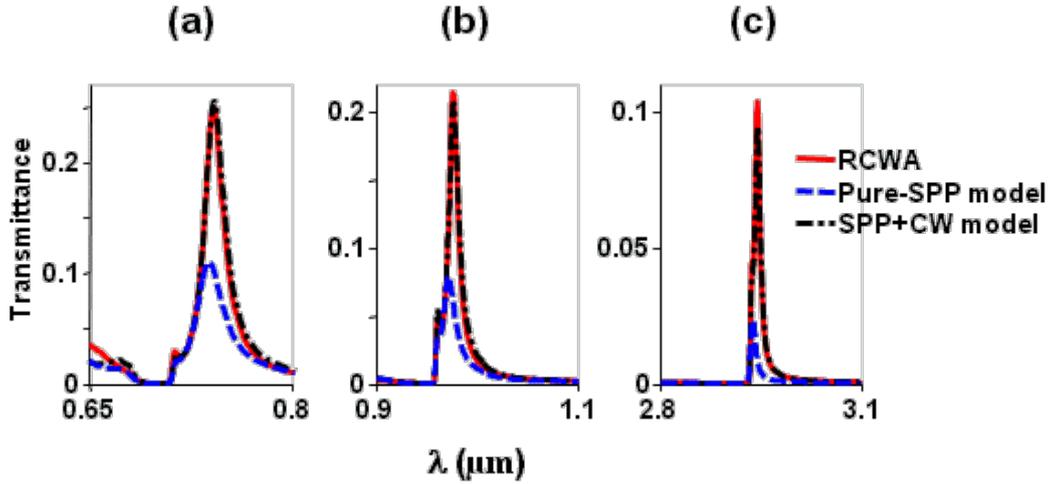

**Fig. 8. The role of SPPs in the EOT.** Three spectral bands are covered, from visible to near-infrared frequencies: **(a)** $a$=0.68 μm, **(b)** $a$=0.94 μm and **(c)** $a$=2.92 μm, $a$ being the grating pitch. The red-solid curves represent fully-vectorial data of the EOT and the blue-dashes are predictions obtained with the pure-SPP model. The black dash-dot curves (almost superimposed with the fully-vectorial results) are obtained with a microscopic model that takes into account SPPs and quasi-CWs (see Section 6.2). The data are obtained for a gold membrane in air perforated by a periodic array of square holes illuminated by a normally incident plane wave. The hole side length is 0.28$a$ (hole filling fraction 8%) and the membrane thickness is $d$=0.21$a$.

## 6. Generalized microscopic model with surface plasmon polaritons and quasi-cylindrical waves

The pure-SPP microscopic model captures most of the important features of the EOT at visible frequencies, but it is largely inaccurate at longer wavelengths. In this Section, we keep on elaborating on an intuitive microscopic description of multiple scattering phenomena occurring on metallic subwavelength surfaces. By incorporating the quasi-CW into the pure SPP model, we obtain a more accurate model that provides quantitative predictions well above the visible wavelengths. In the first subsection 6.1, a key scattering process of surface waves, the cross conversion from quasi-CWs to the SPPs is investigated. In the second subsection 6.2, the quasi-CW contribution is incorporated into the pure-SPP model, and a generalized wavy model, similar to the pure-SPP model (actually it is much more accurate as shown by the black dash-dot curves in Fig. 8), is presented.

### 6.1 Cross conversion from quasi-cylindrical wave to surface plasmon polariton

For a metal surface patterned with a set of 1D indentations under external illumination by TM-polarized light, several scattering processes of surface waves may exist: the SPP-to-SPP scattering that has been considered in the pure-SPP description, the possible CW-to-SPP or SPP-to-CW cross conversions, and the CW-to-CW scattering. The cross conversion between different surface waves plays a key role in the physical multiple-scattering picture. Demonstration of its existence along with a quantitative description of its scattering coefficient appear to be a heuristic step in incorporating quasi-CWs to build up an accurate microscopic description of subwavelength metallic surfaces.

The importance of the cross conversion has been demonstrated in [Yan09], by considering a groove doublet and by calculating its SPP excitation efficiency on the outer sides as a function of the groove separation-distance. The interpretation of the computational results has led the authors to conclude that the SPP excitation efficiency dependence on the separation distance cannot be explained if one does not consider a CW-to-SPP cross conversion. Additionally, the authors have proposed a method to directly extract, from the SPP excitation efficiency, the scattering coefficients associated to the cross-conversion process and have argued and verified that the cross-conversion scattering coefficients are simply related to SPP scattering coefficients,

$$\rho_c \approx \rho_{SP}, \tag{3a}$$

$$\tau_c \approx \tau_{SP}-1, \tag{3b}$$

where $\tau_c$ and $\rho_c$ are the cross-conversion coefficients from an incident CW to a transmitted and a reflected SPP (Fig. 9a), and $\tau_{SP}$ and $\rho_{SP}$ are the elastic transmission and reflection coefficients of the SPP (Fig. 9b). Equations (3a) and (3b) originate from a map of the scattering of an incident CW to the scattering of an incident SPP by a subwavelength indentation. Although the two incident fields are different in nature, their distributions are similar within the subwavelength region of the indentation, which yields an equality between the two scattered fields with the use of the causality principle. Note that in Eq. (3b), $\tau_{SP}-1$ represents the transmitted SPP amplitude that is scattered by the indentation.

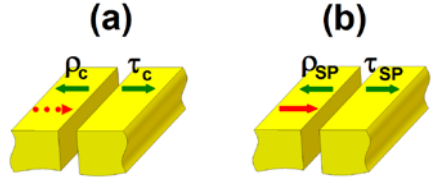

**Fig. 9. Mapping from the cross conversion to the SPP scattering for a slit. (a)** Cross conversion coefficients $\rho_c$ and $\tau_c$ from an incident CW (red dotted arrow) to a reflected and a transmitted SPP (green solid arrows). **(b)** SPP scattering coefficients $\rho_{SP}$ and $\tau_{SP}$ from an incident SPP (red solid arrow) to a reflected and a transmitted SPP (green solid arrows).

## 6.2 Multiple-scattering model with surface plasmon polaritons and quasi-cylindrical waves

In addition to providing closed form expressions, the main force of the pure-SPP model is to propose an intuitive and physical wavy description of the multiple scattering processes involved at metallic subwavelength interfaces. In order to make the model more accurate, one should introduce the quasi-CW into the pure-SPP formalism, and define scattering coefficients for quasi-CWs, including the CW-to-CW scattering and the cross conversion as discussed in the previous subsection.

To derive a generalized formalism, it is convenient to introduce the concept of hybrid waves (HWs) [Liu10]. For 1D subwavelength indentations, the generalized wavy formalism relies on two main ingredients. The first ingredient is related to the overall shape of the field scattered by subwavelength indentations on metallic surfaces. This shape is always composed of a known mixing ratio of SPP and quasi-CW waves at a given frequency, and the respective contributions are fixed, independently of the excitation field and of the exact geometry of the indentation (provided that the indentation is subwavelength, indeed). The property is illustrated in Fig. 10, which shows the fields scattered on the metal interface for several subwavelength indentations and for various incident illuminations. The fields are calculated with a fully vectorial method and are normalized so that their amplitude are all equal at a distance $|x| = \lambda$ from the indentation. Remarkably, it is found that for every frequency, all the scattered fields are identical, except for a proportionality factor.

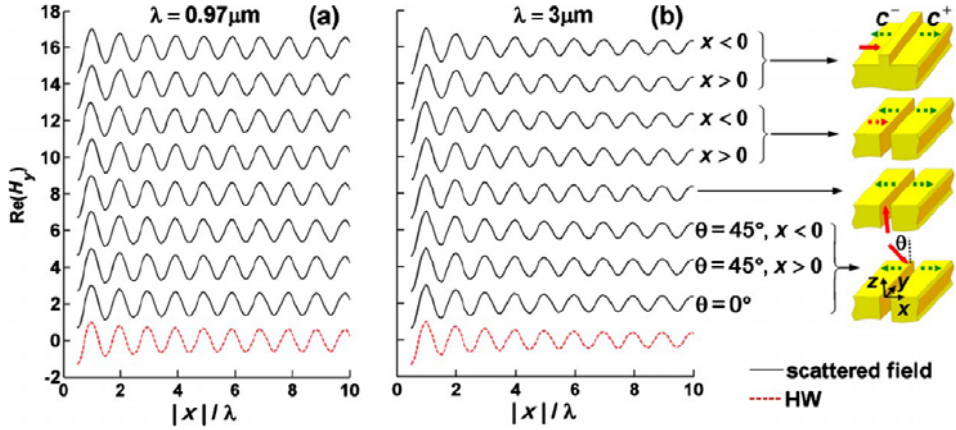

**Fig. 10 Field scattered by a single 1D subwavelength indentation on a metallic surface under TM-polarized illumination (from [Liu10]).** (a)-(b) Magnetic fields $H_y$ scattered on the surface (at $z$=0) for $\lambda$=0.97 and 3 µm. They are vertically shifted by 2 and normalized such that $H_y(|x|=\lambda)$=1. The black solid curves are data calculated with the fully vectorial method and the two red dashed curves show the HW calculated as the radiation of a $y$-polarized magnetic line source on the surface. The results are gathered for gold slits and ridges with widths 0.27$\lambda$ and ridge height 0.27$\lambda$. From bottom to top, the illuminations are a plane wave, the fundamental slit $TEM_{00}$ mode, a HW generated by a magnetic line source located on the surface at $x=-\lambda$, and a SPP mode. The dashed and solid arrows on the surface represent HWs and SPPs, respectively, the arrows in the slit represent fundamental slit modes, and other arrows in free space represent plane waves. The arrows denoting incident and scattered waves are in red and in green, respectively. This notation of arrows is consistently used throughout the chapter.

This important property comes from the fact that under TM-polarized illumination, any 1D subwavelength indentation can be approximated by two coherent electric line sources, one with a polarization parallel to the surface and the other with a polarization perpendicular, and that the radiations of these two sources are *approximately* equal (they are strictly equal in the limit of large metal conductivity). This has been shown in [Lal09] by an analytical treatment. In particular, the scattered fields are composed of a SPP and of a quasi-CW with a fixed mixing ratio, and thus this mix forms a new wave with universal properties [we call a hybrid wave (HW) hereafter]. Note that the HW is nothing else than the Green function of a metal-dielectric interface for a dipole line source on the interface. The only new point we stress here is that the Green function is *almost* independent of the line source polarization (it is "degenerate").

Therefore, in the multiple scattering processes of any subwavelength metallic surfaces,

SPPs and quasi-CWs only appear in a fixed proportion at a given frequency, or in another word, only HWs exist on the surface. This property largely simplifies the introduction of the quasi-CW into the pure-SPP model, since the four scattering processes, the SPP-to-SPP, CW-to-SPP, SPP-to-CW and CW-to-CW, may be combined into a single HW-to-HW scattering process.

The second ingredient of the generalized wavy formalism is the definition of scattering coefficients for HW. Although HWs are not normal modes (just like quasi-CWs, they are not exponentially damped, they do not possess phase or group velocities …), it can be shown that it is possible to define scattering coefficients for HWs, in the same way as scattering coefficients have been defined for the bound SPP modes in Section 5 (see Fig. 7), and that the scattering coefficients are equal to those of the SPP. For instance, if we refer to the HW scattering coefficients in Fig. 11 for a semi-infinite slit, all the HW scattering coefficients can be related to the classical SPP scattering coefficients, and as shown in [Liu10], we may write

$$\beta_{HW}(k_x) = \beta_{SP}(k_x), \ \alpha_{HW} = \alpha_{SP}, \tag{4a}$$

$$\beta'_{HW}(k'_x) = \beta'_{SP}(k'_x), \ \alpha'_{HW} = \alpha'_{SP}, \tag{4b}$$

$$\rho_{HW} = \rho_{SP}, \ \tau_{HW} = \tau_{SP} - 1, \tag{4c}$$

where the subscripts HW and SP refer to HW and SPP, respectively. Equations (4a-4c) are remarkably simple and readily relate non-intuitive HW scattering coefficients to much classical SPP coefficients that are routinely calculated with various numerical tools. Additionally, they allow us to preserve the intuitive picture of a microscopic wave progression, and to explicitly analyze the macroscopic properties of metallic surface in terms of a multiple scattering process. The equalities between the HW scattering coefficients and their associated SPP ones are justified in [Liu10]. Although the HW scattering coefficients may be directly extracted from the calculated scattered field, this calculation cannot benefit from classical normal-mode theory [Vas91] since the HW is not a mode. Equations (4a-4c) render the calculations of the HW scattering coefficients much simpler, since the coefficients can be obtained directly from the scattering coefficients of SPPs, and therefore reciprocity arguments (under proper normalization [Liu10]) may be applied even if the HWs are not normal modes.

From the elementary HW scattering coefficients, it is easy to derive a coupled-wave model that provides closed-form expressions for the transmittance and reflectance (thus absorbance) of subwavelength metallic surfaces. In [Liu10], the model has been tested for various geometries such as grooves and ridges, or mix of grooves and ridges. In all cases, comparisons with fully vectorial computational results have revealed that the generalized formalism is highly accurate, even when the indentation dimensions are as large as λ/3. Similar results have been reported in [Hua11, Li11] for other geometries. The generalized formalism has also been successfully applied for the EOT, see [Liu10] for details. In the generalized formalism, the reflection coefficient $r_A$ of the fundamental supermode of the hole array is given by

$$r_A = r + \frac{2\alpha_{SP}^2}{(1/\Sigma H_{HW} + 1) - (\rho_{SP} + \tau_{SP})}. \tag{5}$$

Actually Eq. (5) is very similar to that obtained with the pure-SPP model, see Eq. (2), except that the SPP phase-term $u^{-1}$ is replaced by $(1/\Sigma H_{HW}+1)$, where $\Sigma H_{HW}$ is a lattice summation of the HW fields that is known analytically [Liu10]. As shown with the black dash-dot curves in Fig. 8, the HW model accurately predicts the EOT from visible to middle-infrared bands. Other computations have shown that the reflectance and the absorbance are also predicted with a high accuracy. It is important to realize that the HW model does not require additional computations, in comparison to the pure-SPP model. Importantly, it relies on the same SPP scattering coefficients, which are therefore found to play a fundamental role in the electromagnetic properties of subwavelength metallic surfaces.

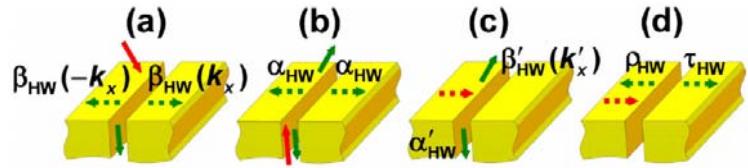

**Fig. 11. HW Scattering coefficients for a slit.** (**a**)-(**b**) Scattering coefficients $\beta_{HW}(k_x)$ or $\alpha_{HW}$ corresponding to HW excitations under illumination either by a TM-polarized incident plane wave with an in-plane parallel wave vector $k_x$ or by the fundamental slit mode. (**c**) Reciprocal scattering coefficients $\beta'_{HW}(k'_x)$ and $\alpha'_{HW}$ under illumination by HWs, where $k'_x$ denotes in-plane parallel wave vectors of scattered plane waves. (**d**) In-plane scattering coefficients $\tau_{HW}$ and $\rho_{HW}$ that characterize the transmission and the reflection of HWs by the slit.

## 7.Conclusion

Many optical phenomena related to subwavelength metallic surfaces, which are observed with metallic nanostructures at visible frequencies, can be "reproduced" at longer wavelengths by scaling the geometrical parameters. At an elementary level, these phenomena are due to the electromagnetic fields that are scattered by the indentations and that interact with the neighbor indentations. For visible wavelengths, the analysis promotes an interaction mediated by surface–plasmon–polaritons (SPPs) and supplemented at distances up to a few wavelengths by an additional scattered near–field, the quasi–cylindrical wave (quasi-CW). At longer wavelength, because they spread far away into the dielectric medium, the delocalized SPPs are marginally excited by the indentations and the quasi-CWs are dominant (Section 4).

The two–wave picture represents a helpful microscopic view to comprehend the rich physics of subwavelength metallic surfaces. The SPP and quasi-CW scattering involves SPP-to-SPP and CW-to-CW scatterings, CW–to–SPP and SPP-to-CW cross conversions, and scattering into radiation modes. All those scattering coefficients, some of them being non trivial, are quantitatively equal to SPP-scattering coefficients (Section 6). This places SPP scatterings at the root of the physics of subwavelength metallic surfaces, even when quasi-CWs are dominantly excited like in the infrared. The important fact that quasi-CWs and SPPs essentially scatter identically is at the core of the concept of hybrid–waves (Section 6.2).

After 100 years or more, research in the area of subwavelength metallic surfaces and gratings continues unabated. This reflects the underlying importance of metal to manipulate light. It is likely that this situation will continue. The use of interfaces possessing complex subwavelength textures is really only beginning and the microscopic point of view presented here may help to understand and to design the surfaces. It is hoped that this review will stimulate new ideas and lead to new research.

## *Acknowlegements*

Haitao Liu acknowledges financial supports from the National Natural Science Foundation of China (No. 10804057), from the Cultivation Fund of the Key Scientific and Technical Innovation Project, Ministry of Education of China (No. 708021), from the 973 Project (No. 2007CB307001), and from the Natural Science Foundation of Tianjin (No. 11JCZDJC15400). Jean Claude Rodier, Lionel Aigouy, Xiaoyan Yang, Jacques Giérak, Eric Bourhis, Christophe Sauvan, Stéphane Collin, Lionel Jacobowiez and Jean Paul Hugonin are acknowledged for fruitful discussions.